# A Comparative Review of Microservices and Monolithic Architectures

Omar Al-Debagy
*Department of Electronics Technology*
*Budapest University of Technology and Economics*
Budapest, Hungary
omeraldebagy@gmail.com

Peter Martinek
*Department of Electronics Technology*
*Budapest University of Technology and Economics*
Budapest, Hungary
martinek@ett.bme.hu

***Abstract*— Microservices' architecture is getting attention in the academic community and the industry, and mostly is compared with monolithic architecture. Plenty of the results of these research papers contradict each other regarding the performance of these architectures. Therefore, these two architectures are compared in this paper, and some specific configurations of microservices' applications are evaluated as well in the term of service discovery. Monolithic architecture in concurrency testing showed better performance in throughput by 6% when compared to microservices architecture. The load testing scenario did not present significant difference between the two architectures. Furthermore, a third test comparing microservices applications built with different service discovery technologies such as Consul and Eureka showed that applications with Consul presented better results in terms of throughput.**

***Keywords — Microservices Architecture, Monolithic Architecture, Performance Evaluation***

## I. Introduction

Nowadays, many companies, such as Netflix, Amazon, and eBay, have migrated their applications and systems to the cloud, because cloud computing model allows these companies to scale their computing resources as per their usage [1]. Martin Fowler defined Microservices Architecture as an approach of developing a suite of small services working as a single application. The services are communicating through lightweight mechanisms, such as an HTTP resource API and each service is running independently in its own process [2].

On the other hand, monolithic architecture is an application with a single code base that includes multiple services. These services communicate with external systems or consumers via different interfaces like Web services, HTML pages, or REST API [3].

Chen et al. claim that Microservices architecture will ease the processes of maintainability, reusability, scalability, availability, and automated deployment when it will be utilized, and these are considered the advantages of microservices architecture. [1].

The main advantages of microservices architecture are the followings. First, microservices can rely on technology heterogeneity, which means each service in one system can use different technology than the other services to achieve the desired goals and performance [4]. Second, another advantage of microservices is if one component of the system fails then it does not affect the whole system. Newman called this advantage as resilience in his book entitled Building Microservices [4]. The third advantage is that the process of scaling can be more accessible compared to monolithic application scaling because only the services that need actual scaling are scaled in the microservices architecture, contrary to a monolithic application requires to be scaled as a whole unit which may lead to higher hardware usage [4]. Fourth, ease of deployment, because with microservices each service can be deployed independently without affecting the performance of other services. Fifth, microservices architecture helps companies to align its architecture with its organizational structure, which will help them to minimize the number of people that are working on a specific codebase. Consequently, microservices enables the organizational alignment [4]. Further advantages are composability and optimizing for replaceability [4].

On the other hand, in the monolithic architecture, applications can be created of tens or hundreds of different services that are tightly coupled in a monolith codebase. This can create plenty of difficulties for teams working in the same environment. Therefore, many companies are moving toward microservices architecture to enable their development teams to coordinate with each other easily [5].

Consequently, in this paper there is a comparison between microservices and monolithic architectures in term of performance, to determine how these architectures perform in different scenarios utilizing different testing setups. Also, because of some contradiction in the literature available when compared the performance of these architectures, led this paper to investigate more about the performance of microservices and monolithic architectures.

## II. Literature Review

This part of the paper presents selected literature discussing microservices and comparing its performance to other architecture such as monolithic and service-oriented architectures. Other researchers' work is also included containing fundamental concepts and ideas about microservices' architecture.

In a research that was done by Singh and Peddoju, the performance of a monolithic application is compared to a microservices application that they developed and compared the performance of the two applications, their tests consisted of 2000 threads. Their results exhibited that microservices architecture has a better performance in terms of throughput when it is used for a large number of requests [5].

IBM research team tackled the same idea [6], but they concentrated on the resource consumption analysis and they compared the performance of the monolithic and microservices applications in different environments and configurations such as in the case of different numbers of CPU cores. Their results showed a significant performance boost in monolithic architecture applications in many configurations and environments, which in a way contradicts the results shown by Singh and Peddoju [5]. Thus, these contradictions will lead this paper to investigate further into the performance differences between these architectures.

Microservices are often compared to Service Oriented Architecture as Mark Richards did in his book. The comparison of microservices' architecture with Service Oriented architecture in term of service and architecture characteristics as well as architecture capabilities are presented in [7].

Villamizar et al. compared the costs of using the cloud to run web applications with different architectures such as microservices, monolithic and Amazon Web Services Lambda architectures. In addition to cost comparison, they also compared response time of each application that they created, which showed that response time increased when microservices architecture was utilized compared to the monolithic architecture because each request must go through the gateway to every microservice in the system [3].

In another paper, the performance of microservices in container-based and virtual machine (VM)-based environment was compared. Amazon cloud environment was applied to conduct their experiments, and they compared the performance of these environments regarding throughput, response time, and CPU consumption. This paper concludes that VM-based environments on Amazon cloud services outperformed container-based environments on Amazon cloud environment, especially concerning response time where VM-based environment showed a better performance of 125% over container-based environment [8].

Docker is a widely used container-based virtualization software that is utilized in microservices architecture and uses Linux containers for the operating system virtualization [9]. The main reason behind using Docker for microservices is the minimal impact of its imposes on processing, memory, and network [10].

Service discovery is an essential component of any microservices application because the location of a microservice is not assigned at the design stage. Also, it may be deployed in a cloud-based environment which means services could relocate and replicate at production systems [11].

## III. Methodology

In order to compare the performance of two different architectures, first of all, there should be an application that can produce the results of microservices and monolithic applications, which can be compared and evaluated. This paper presents results created by a development platform known as JHipster utilized to generate web applications that consist of Spring Boot and Angular JS frameworks. The application that was developed for this particular paper consisted of three services.

A typical JHipster application will include three components. First, JHipster Registry is an essential component in the microservices architecture because it connects all the other components with each other and provides the communication between these components. Second, the microservice application which will provide the backend capabilities through exposing the API. Third, the microservices gateway is the frontend of the whole system which will include all the APIs of every microservice application in the system [12].

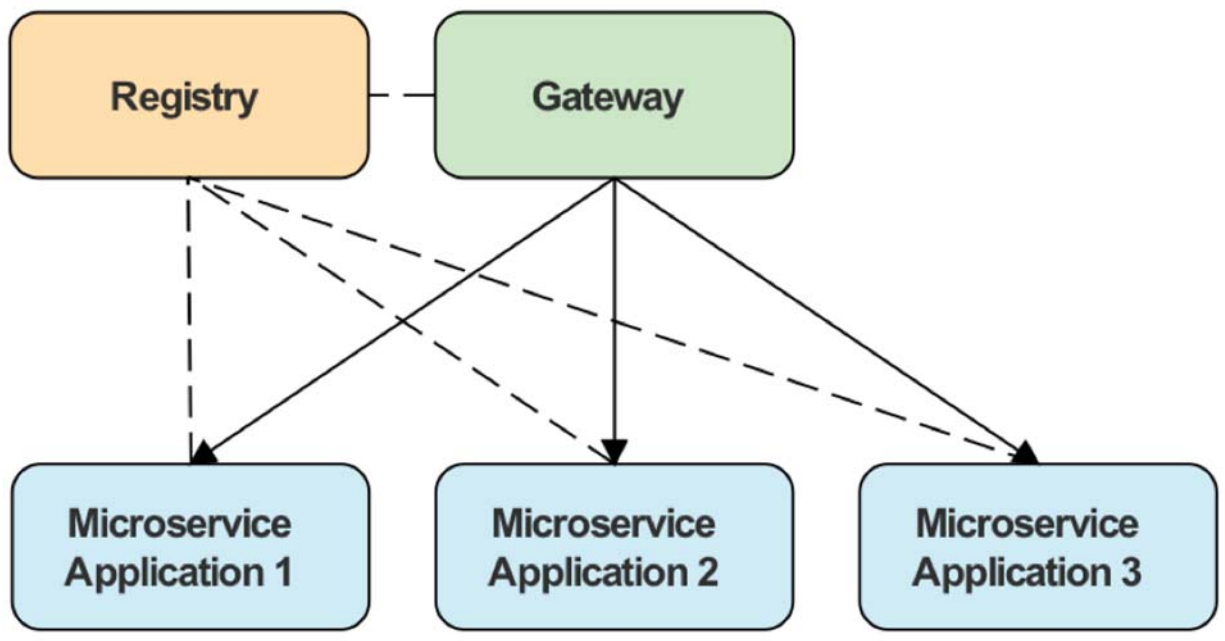

Fig. 1: General overview of microservices architecture

JMeter [13] was used to test the performance of these applications. Two test scenarios were set up to compare the performance of the microservices application and monolithic application. Also, another test scenario was created to compare the effect of different technologies on the performance of the microservices application.

Response time and throughput were utilized to compare the performance of the tested applications. Response time is the time it takes for a client to receive a response from the server for a request of a specific service. In other words, "Response Time is the time elapsed between the request and reply" [14].

Another performance metric is throughput, which is the number of requests an application can handle per second. Therefore, it means dividing the total number of processed requests by the time it took to process all the requests.

The first test scenario is load testing which is used to monitor the effects of increasing the number of users on the application and how it will affect throughput and response time. It starts with 100 threads with a ramp-up of 2 minutes and holds time of another 2 minutes to analyze the concurrency of the system, then increases the number of threads until 7000 threads each time with 2 minutes for ramp-up and holds time.

The second test scenario is concurrency testing which is used to check how the system will hold up if all the services are used at the same time, so the test was designed to send requests to each service through their exposed APIs at the same time. It started with 100 requests for each service with no specific ramp-up time and increasing the number of requests gradually until 1000 requests.

The third testing scenario tested the endurance of the system which consisted of 10000 threads with a 10 minutes ramp up and a 10 minutes hold time, but this time the test included other configurations of microservices architecture using different technologies for the service discovery such as Consul and Eureka.

JMeter was installed on a remote client and connected to the server via ethernet cable to ensure the reliability of the network. Docker was utilized to run the applications on containers, and the server that was used to run Docker environment had 16GB of memory and 2.60 GHz of CPU.

## IV. Results and Discussion

After running the test scenarios for each scenario and collecting the results of these tests, each architecture and test showed different results in terms of throughput and response time.

### A. *First Test Scenario: Load Testing*

This test compared the performance of microservices architecture and monolithic architecture in terms of throughput, response time, and how many requests each application can handle throughout the testing period. In this test, the ramp-up time was 2 minutes and the hold time was 2 minutes as well. The number of threads was set to 100 and was increased gradually until 7000.

The results presented a similar performance between microservices and monolithic architectures. It is apparent in Fig. 2, which shows the average throughput of all five tests that are done in the first scenario. The monolithic application throughput is better at the test with 100 threads which leads to the indication that monolithic architecture can perform better with a small number of users because microservices architecture needs to use communication interface between each service and its database. Then the performance of each architecture will decrease while the number of threads increases. This happened because the increased number of users increases the response time of requests which leads to the decline of the overall throughput. Although, the performance of the monolithic application is better with 100 and 1000 threads eventually it will be similar to the performance of the microservices application with 6000 and 7000 threads. Based on this test scenario, it is evident that the monolithic application performs much better with a small number of users compared to the microservices application. The difference in performance between the monolithic application and microservices application was 0.87% on average which did not show any significant difference between the two approaches in this test scenario.

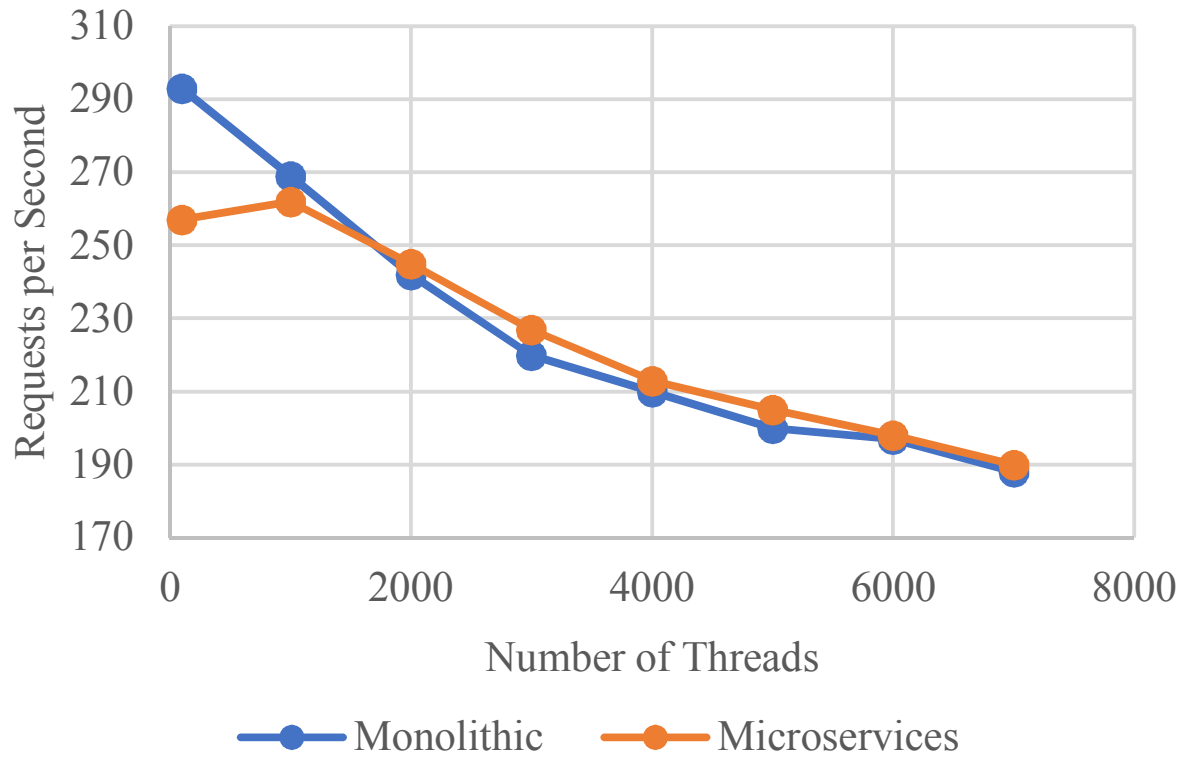

Fig. 2: Throughput of first test scenario

Another test metric obtained from this test scenario was response time. There was not a big difference in response time comparing microservices and monolithic architecture. The response time increased by increasing the number of threads as expected.

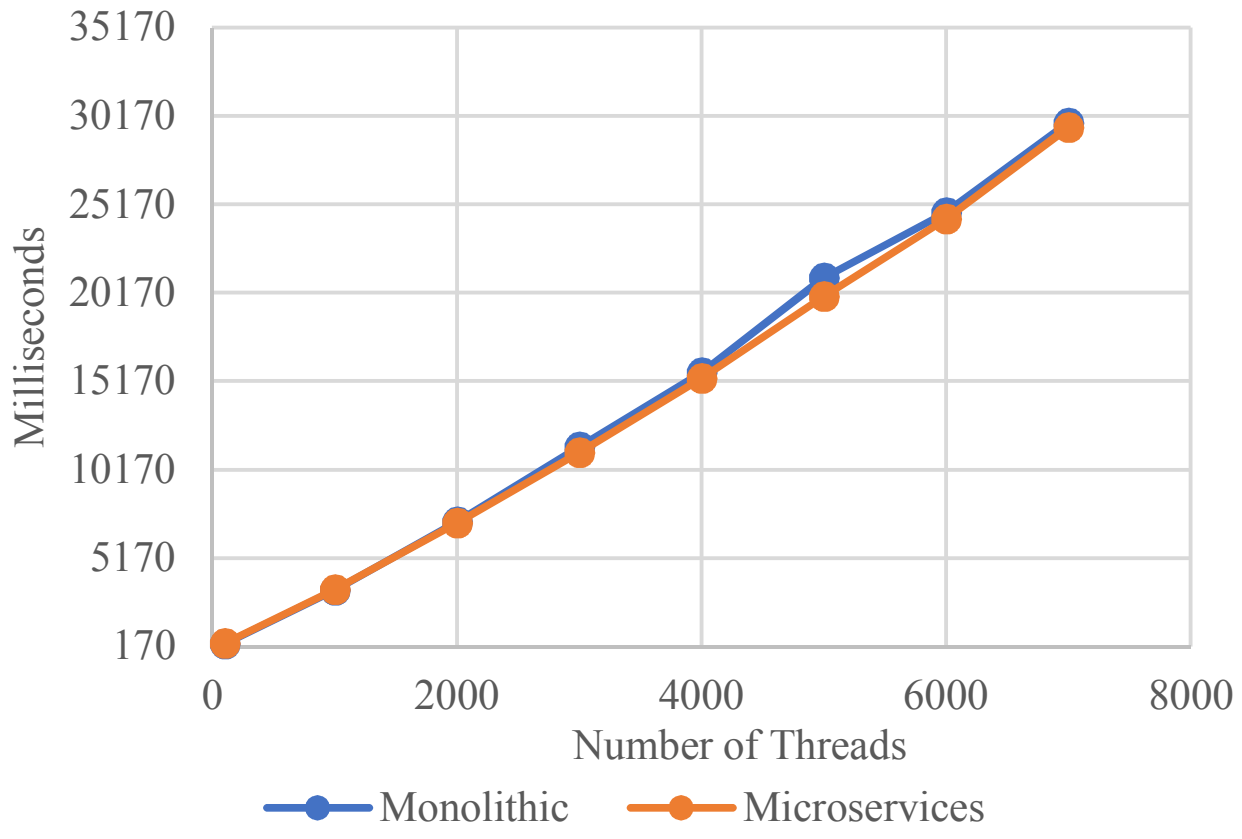

Fig. 3: Response Time of first test scenario

The number of fulfilled requests is the third metric chosen to compare the performance of microservices and monolithic architectures. The result of this test is a little bit similar to the average throughput case presented in Fig. 2, but this metric could show the difference between the two architectures in more details. As it is the same case with the throughput performance, the monolithic application has the upper hand when it comes to the small number of requests. Although the number of threads increases, the number of fulfilled requests decreases at both architectures, as it is apparent in Fig. 4 that microservices architecture can fulfill more requests compared to monolithic architecture.

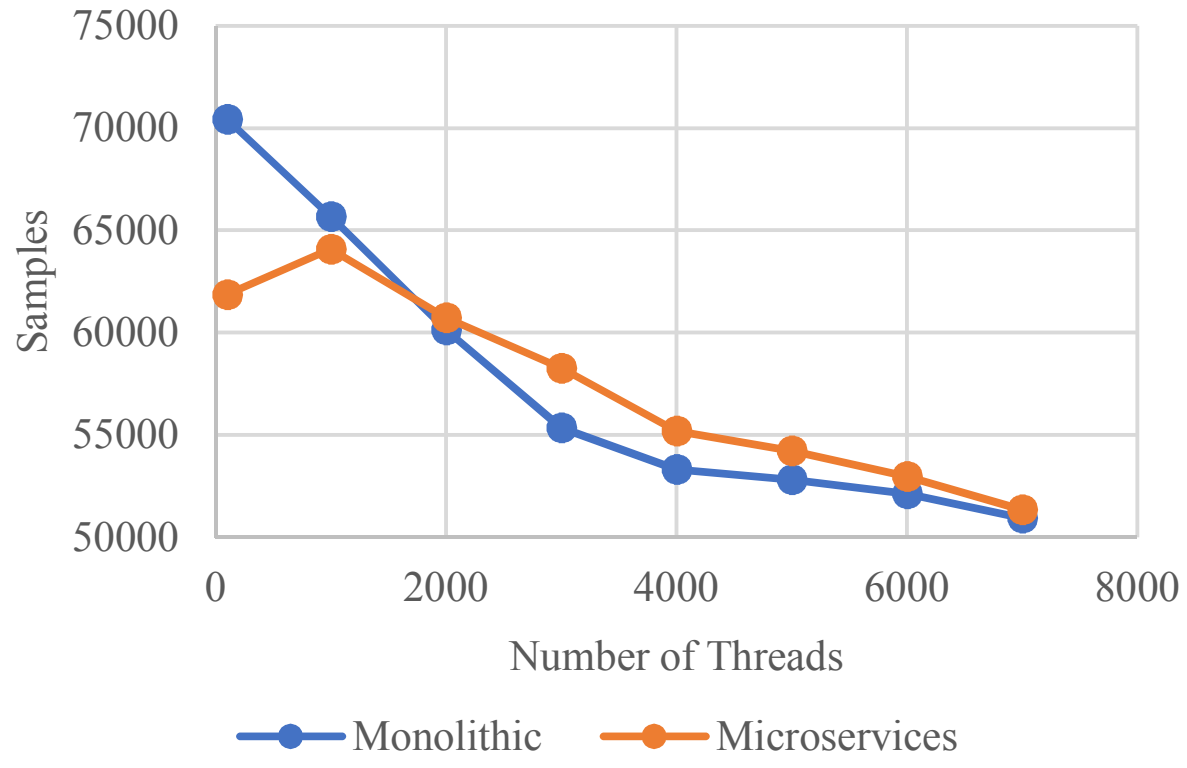

Fig. 4: Number of processed requests for first test scenario

### B. *Second Test Scenario: Concurrency Testing*

The second test was created to evaluate the performance of the architecture at a higher load than in the first test. All the services were invoked simultaneously without setting any specific ramp-up in order to make the threads run at once without any wait time. So, the results showed that monolithic architecture performed better than the microservices architecture regarding throughput or how many requests the application could handle per second. Monolithic architecture showed better performance in terms of throughput by 6% on average, when comparing the performance of all the trials of the second testing scenario. This shows a contradiction to the results that were presented by Singh and Peddoju [5] but similar to the results of IBM research team [6].

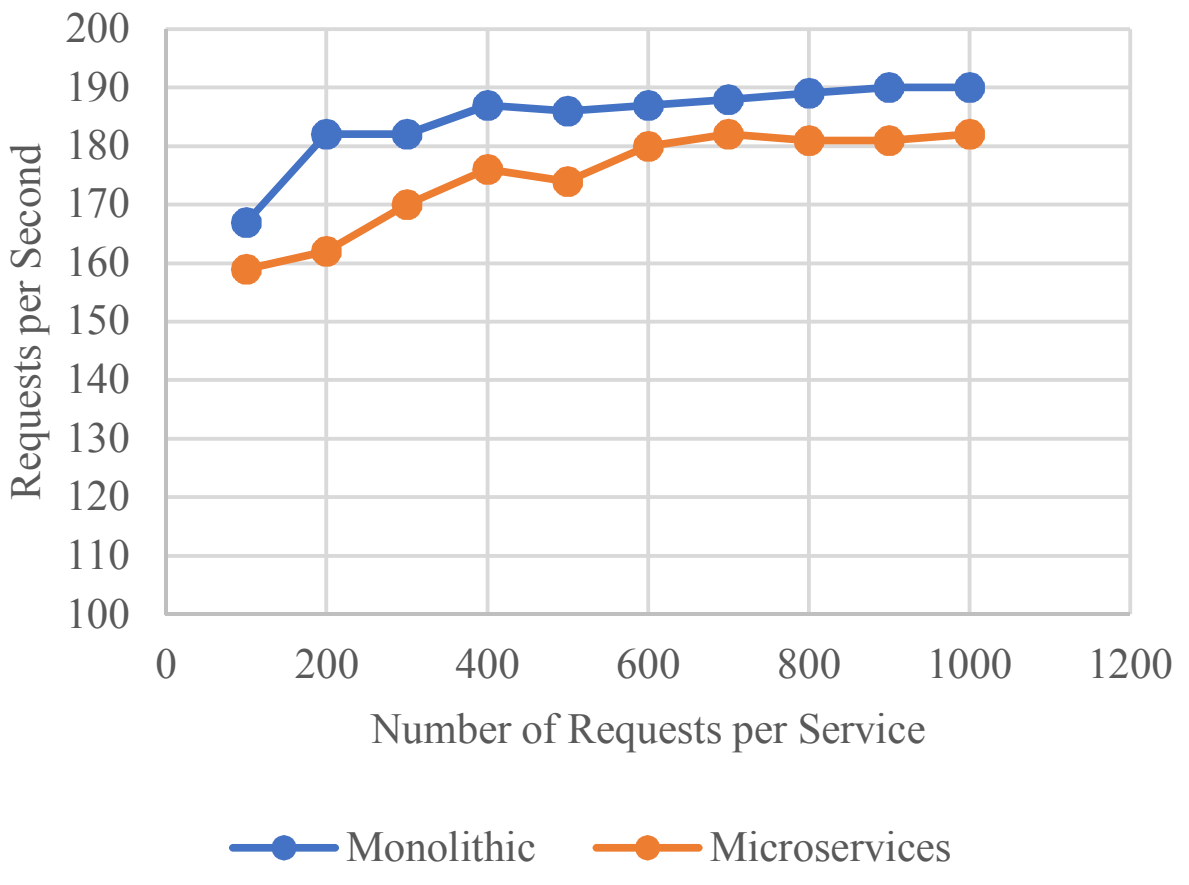

Fig. 5: Throughput for second test scenario

However, when it comes to the response time, there was no significant difference between the two architectures. Both of them started around 1790 milliseconds with 100 threads, and then it increased until it reached around 15000 milliseconds with 1000 threads. This test also ran for three times in order to make sure that the results were reliable.

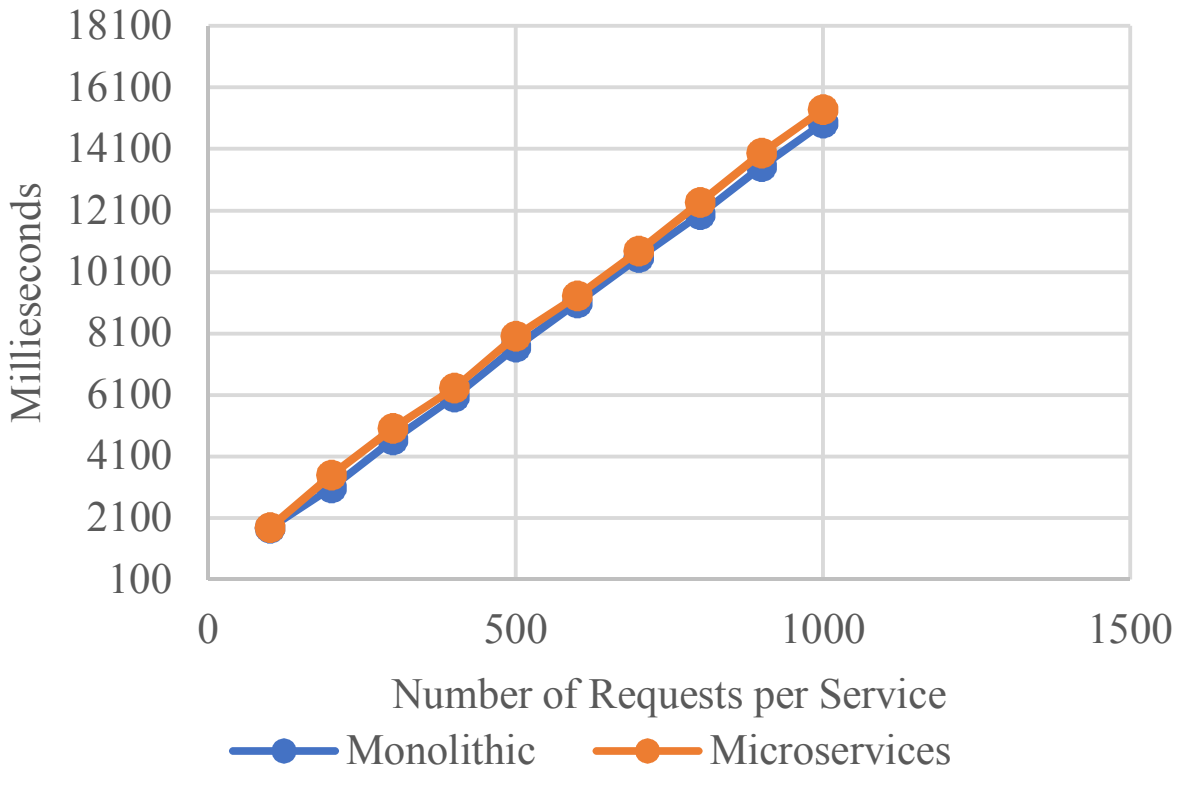

Fig. 6: Response Time for second test scenario

### C. *Third Test Scenario: Eureka vs Consul Service Discovery Test*

This scenario used 10,000 threads through 20 minutes divided as 10 minutes ramp up with 5 step ups, which means every 2 minutes the number of threads was increased by 2000 until it reached 10,000 threads. The hold time was set to 10 minutes after that. In this scenario, a new services discovery technique was used in the microservices architecture which is called Consul, and it was compared to the default services discovery technology that is utilized by JHipster which is the Eureka services discovery. The results displayed that microservices with Consul service discovery had a higher throughput compared to microservices with Eureka by 3.8% difference in throughput performance.

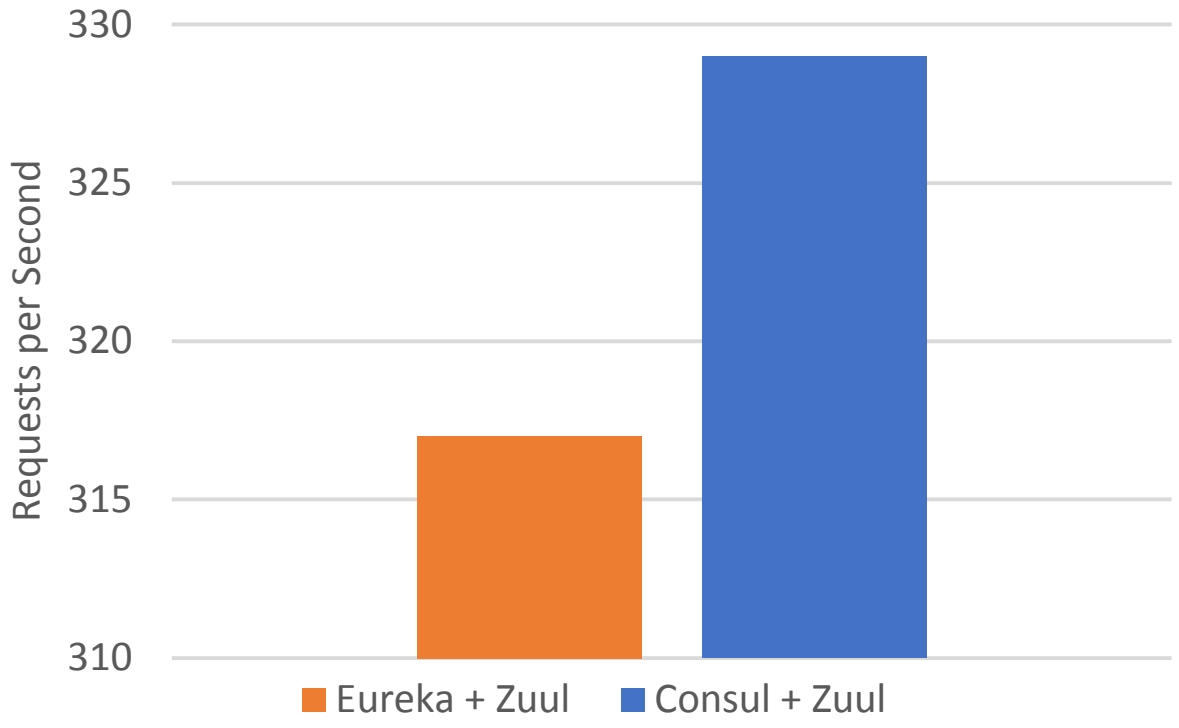

Fig. 7: Throughput for third test scenario

Regarding the average response time of the third test scenario that was done between different configurations of microservices application services, Consul and Zuul configuration showed less response time compared to other configuration such as Eureka with Zuul. For example, the average response time of Consul was 23254 milliseconds whereas response time of Eureka configuration was 23841 milliseconds, which shows that the microservices application with Consul service discovery had a little better response time than the microservices application with Eureka service discovery.

However, when it comes to response time, there was no significant difference between the two architectures. Both of them started around 1790 milliseconds with 100 threads, and then it increased until it reached around 15000 milliseconds with 1000 threads. This test was repeated three times in order to make sure that the results were reliable.

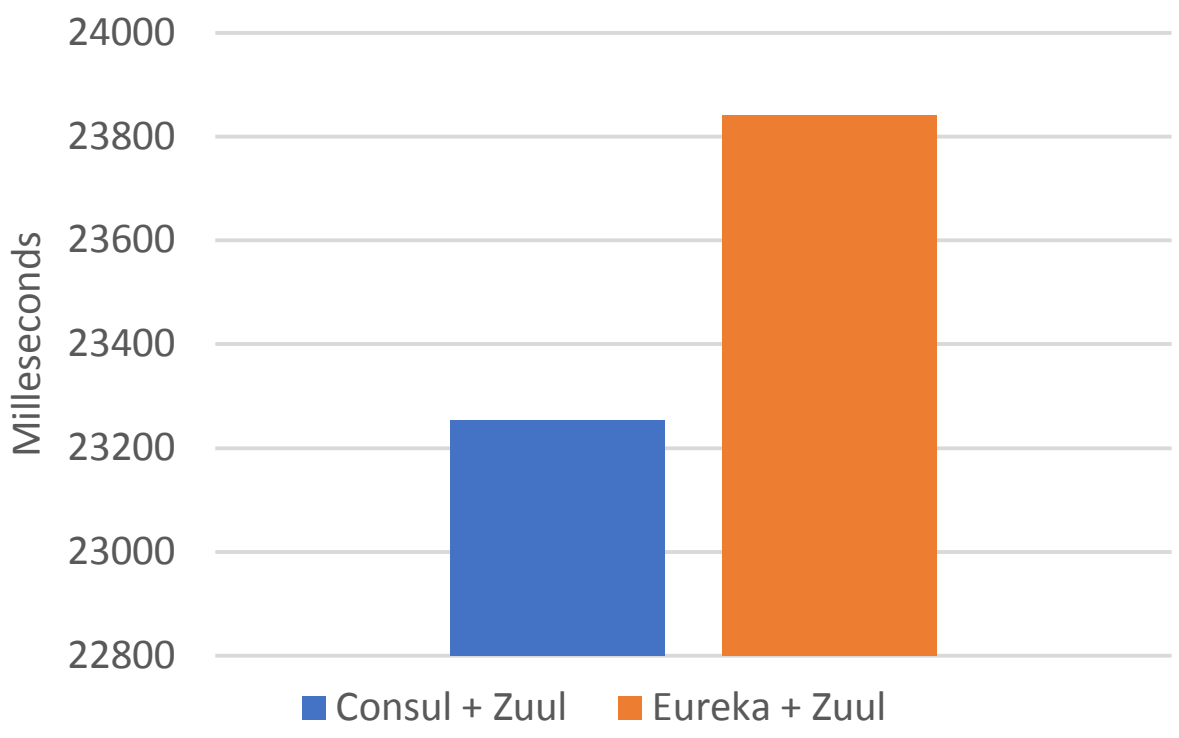

Fig. 8: Average response time of third test scenario

## V. Conclusion

Analyzing the results of the first test scenario this research can conclude that microservices and monolithic application can have similar performance under normal load on the application. In the case of a small load with less than 100 users, the monolithic application can perform a little bit better than microservices application. Hence monolithic application is recommended for small applications used only by a few users. In the second test scenario, the results were different in term of the throughput. The number of requests was fixed so that to find the exact number of requests an application can handle per second. The monolithic application showed higher throughput on average. Thus, the monolithic application can handle requests in a faster manner, so the monolithic application can be used when the developer especially aims that the application handles requests in a faster way. Another test scenario included a comparison between two microservices application with different service discovery technologies such as Eureka and Consul. The results of this test indicated that the microservices application with Consul service discovery performed better than the application with Eureka service discovery technology in terms of throughput or the number of handled requests per second which displayed an improvement by 4% when using Consul. Therefore, microservices application with Consul as a service discovery technology can be preferred compared to microservices with Eureka services discovery technology.